\documentclass[12pt,A4]{article}
\usepackage{amssymb,amsfonts,amsthm,amsmath}

\usepackage{epsf}
\usepackage{graphicx}
\usepackage{placeins}
\usepackage{setspace}

\def\hang{\hangindent\parindent}
 \def\rf{\par\noindent\hang}

\begin{document}
\baselineskip=21pt

\begin{center} \Large{{\bf
Confidence Intervals in Regression Utilizing Prior Information}}
\end{center}

\bigskip

\begin{center}
\large{
{\bf Paul Kabaila$^*$, Khageswor Giri}}
\end{center}

\begin{center}
{\sl Department of Mathematics and Statistics,
La Trobe University, Victoria 3086, Australia}
\end{center}

\medskip

\noindent {\bf Abstract}

\medskip

We consider a linear regression model with regression parameter $\beta = (\beta_1, \ldots,  \beta_p)$ and
independent and identically $N(0, \sigma^2)$ distributed errors.
Suppose that the parameter of interest is $\theta = a^T \beta$ where $a$ is a specified vector.
Define the parameter $\tau = c^T \beta - t$ where the vector $c$ and the number $t$
are specified and $a$ and $c$ are linearly independent.
Also suppose that we have uncertain prior information that $\tau = 0$.
We present a new frequentist $1-\alpha$ confidence interval for $\theta$ that utilizes this
prior information. We require this confidence interval to (a) have endpoints that are
continuous functions of the data and (b)
coincide with the standard $1-\alpha$ confidence interval
when the data strongly contradicts this prior information.
This interval is optimal in the sense that it has minimum
weighted average expected length where the largest weight is given to this expected
length when $\tau = 0$. This minimization leads to an interval that has the following desirable properties.
This interval has expected length that (a) is relatively small when the prior
information about $\tau$ is correct and (b) has a
maximum value that is not too large.
The following problem will be used to
illustrate the application of this new confidence interval.
Consider a $2 \times 2$ factorial experiment with 20 replicates.
Suppose that the parameter of interest $\theta$ is a specified {\it simple} effect
and that we have uncertain prior information that
the two-factor interaction is zero. Our aim is to find a frequentist
0.95 confidence interval for $\theta$ that utilizes this
prior information.

\bigskip


\vspace{0.7cm}

\noindent $^*$Corresponding author. Tel.: +61 3 9479 2594, fax: +61 3 9479 2466.

\noindent {\sl E-mail address:} P.Kabaila@latrobe.edu.au (Paul Kabaila).

\newpage

\noindent {\large{\bf 1. Introduction}}

\medskip

 Consider the linear regression model $Y = X \beta + \varepsilon$,
 %
 %
 where $Y$ is a random $n$-vector of responses, $X$ is a known $n \times p$ matrix with linearly
independent columns, $\beta = (\beta_1,\ldots, \beta_p)$ is an unknown parameter vector and
$\varepsilon \sim N(0, \sigma^2 I_n)$ where $\sigma^2$ is an unknown positive parameter.
Suppose that the parameter of interest is $\theta = a^T \beta$ where $a$ is specified
$p$-vector ($a \ne 0$). Define the parameter $\tau = c^T \beta - t$ where the vector $c$ and the number $t$
are specified and $a$ and $c$ are linearly independent.
Also suppose that previous experience with similar data sets and/or
expert opinion and scientific background suggest that $\tau = 0$.
In other words, suppose that we have uncertain prior information that $\tau = 0$.
Of course, this includes the particular case that $c=(0,\ldots,0,1)$ and $t=0$, so that the
uncertain prior information is that $\beta_p = 0$.
Our aim is to find a frequentist $1-\alpha$ confidence interval (i.e. a confidence
interval whose coverage probability has infimum $1-\alpha$)
for $\theta$ that utilizes this
prior information, based on an observation of $Y$.

An attempt to incorporate the uncertain prior information that $\tau = 0$
into the construction of a $1-\alpha$ confidence interval for $\theta$ is as follows.
We carry out a preliminary test of the null hypothesis that $\tau = 0$
against the alternative hypothesis that $\tau \ne 0$. If this null hypothesis
is accepted then the confidence interval is constructed assuming that it was known
{\it a priori} that $\tau = 0$; otherwise the standard $1-\alpha$ confidence interval for $\theta$ is
used. We call this the naive $1-\alpha$ confidence interval for $\theta$.
This confidence interval is based on a false assumption and so we expect that its
minimum coverage probability will not necessarily be $1-\alpha$. This minimum coverage
probability has been investigated by Giri and Kabaila (2008), Kabaila (1998, 2005a),
Kabaila and Giri (2009a) and Kabaila and Leeb (2006).
In many cases this minimum is far below $1-\alpha$, showing that this confidence interval
is completely inadequate. So, the naive $1-\alpha$ confidence interval fails to
utilize the prior information that $\tau = 0$.

Whilst the naive $1-\alpha$ confidence interval for $\theta$ fails abysmally to
utilize the prior information that $\tau = 0$, its form (as described in Section 2) will be used to provide
some motivation for the new confidence interval described in Section 3.
Similarly to Hodges and Lehmann (1952), Bickel (1983, 1984), Kabaila (1998), Kabaila (2005b),
Farchione and Kabaila (2008), Kabaila and Tuck (2008) and Kabaila and Giri (2009b),
our aim is to utilize the uncertain prior
information in the frequentist inference of interest,
whilst providing a safeguard in case this prior information happens to be incorrect.
We assess a $1-\alpha$ confidence interval for $\theta$ using the ratio
(expected length of this confidence interval)/(expected length of standard $1-\alpha$ confidence interval).
We call this ratio the scaled expected length of this confidence interval.
In Section 3 we describe a new $1-\alpha$ confidence interval for $\theta$ that utilizes the
prior information. This interval has endpoints that are
continuous functions of the data and it has the following properties.
It coincides with the standard $1-\alpha$ confidence interval
when the data strongly contradicts the prior information.
This interval is optimal in the sense that it has minimum
weighted average expected length where the largest weight is given to this expected
length when $\tau = 0$. This minimization leads to an interval that has the following desirable properties.
This interval has scaled expected length that (a) is smaller than 1 when the prior
information about $\tau$ is correct and (b) has a
maximum value that is not too much larger than 1. The idea of minimizing a weighted average expected
length of a confidence interval, subject to a coverage probability inequality constraint,
appears to have been first used by Pratt (1961).

In Section 4 we consider the following scenario. Suppose that
a $2 \times 2$ factorial experiment, with factors labeled A and B and with
more than 1 replicate, has been conducted. Also suppose that our interest
is solely in the {\it simple} effect of changing factor A
from low to high when factor B is low. Consider, for example, the case that factor A
(B) being low or high corresponds to the absence or presence of treatment A (B), respectively.
Our interest may be solely in the effect of treatment A compared to no treatment
(cf. Hung et al (1995)).
In other words, the parameter of interest $\theta$ is the {\it simple} effect
(expected response when factor A is high and factor B is low) $-$
(expected response when factor A is low and factor B is low).
In this case, $p=4$ and we identify $\tau$ with the two-factor interaction.
Suppose that previous experience with similar data sets and/or
expert opinion and scientific background suggest that the two-factor interaction is zero.
In a $2 \times 2$ factorial clinical trial comparing two drugs whose presumed effects are
on completely different systems and/or diseases, it seems reasonable to suppose
that we have uncertain prior information that the two-factor interaction
is zero (Stampfer et al (1985), Steering Committee of the Physicians' Health Study Research Group
(1988)), Buring and Hennekens (1990) and Hung et al (1995)). For an example of the elicitation of
uncertain prior information in a factorial experiment via expert opinion and scientific background
in a chemical context see Dub\'e et al (1996).

An attempt to utilize the uncertain prior information that the two-factor interaction is zero
is to use a naive $1-\alpha$ confidence interval for $\theta$ constructed using the following preliminary test.
The preliminary test is of the null hypothesis that the two-factor interaction is zero
against the alternative hypothesis that the two-factor interaction is non-zero.
This confidence interval has a minimum coverage probability that is far below $1-\alpha$,
showing that it is completely inadequate. As an illustration, consider the case that the number of replicates is 20,
$1-\alpha = 0.95$ and the preliminary hypothesis test has
level of significance 0.05.
We find, using the methodology of Kabaila (1998, 2005a) or Giri and Kabaila (2008) or Kabaila and Giri (2009a), that
the minimum coverage probability of this confidence interval is 0.7306.
The poor coverage properties
of the naive confidence interval
are presaged by the poor properties of some other inferences carried out
after this preliminary test, see Fabian (1991), Shaffer (1991) and Ng (1994)
(cf.  Neyman (1935), Bohrer and Sheft (1979) and Traxler (1976)).

The properties of the new confidence interval, described in Section 3,
are illustrated in Section 4 by a detailed analysis
of the $2 \times 2$ factorial experiment example with
20 replicates and $1-\alpha = 0.95$.
Define the parameter $\gamma = \tau/\sqrt{\text{var}(\hat \tau)}$,
where $\hat \tau$ denotes the least squares estimator of $\tau$.
As proved in Section 3, the coverage probability of the new confidence interval for $\theta$ is an even function
of $\gamma$. The top panel of Figure 3 is a plot of  the coverage probability of the new 0.95 confidence interval for $\theta$
as a function of $\gamma$. This plot shows that the new 0.95 confidence interval for $\theta$ has
coverage probability 0.95 throughout the
parameter space. As proved in Section 3, the scaled expected length of the new confidence interval for $\theta$ is an even function
of $\gamma$. The bottom panel of Figure 3 is a plot of the square of the scaled expected length of the new 0.95 confidence interval for $\theta$
as a function of $\gamma$. When the prior information is correct (i.e. $\gamma=0$), we gain since the square of the scaled expected length
is substantially smaller than 1. The maximum value of the square of the scaled expected length is not too large.
The new 0.95 confidence interval for $\theta$ coincides with the standard $1-\alpha$ confidence interval
when the data strongly contradicts the prior information. This is reflected in Figure 3 by the fact that the
square of the scaled expected length approaches 1 as $\gamma \rightarrow \infty$.

\bigskip

\noindent {\large{\bf 2. The naive confidence interval}}

\medskip

The naive $1-\alpha$ confidence interval for $\theta$ is constructed as follows.
We carry out a preliminary test of the null hypothesis that $\tau = 0$
against the alternative hypothesis that $\tau \ne 0$. If this null hypothesis
is accepted then the confidence interval is constructed assuming that it was known
{\it a priori} that $\tau = 0$; otherwise the standard $1-\alpha$ confidence interval for $\theta$ is
used. As noted in the introduction, this confidence interval will often have minimum coverage probability
far below $1-\alpha$, showing that it is completely inadequate.
In this section we describe the naive confidence interval in a new form that will be used to provide
some motivation for the new confidence interval described in the next section.

Let $\hat \beta$ denote the least squares estimator of $\beta$. Let
$\hat \Theta$ denote $a^T \hat \beta$ i.e. the least squares estimator of $\theta$.
Also, let $\hat \tau$ denote $c^T \hat \beta - t$ i.e. the least squares estimator of $\tau$.
Define the matrix $V$ to be the covariance matrix of $(\hat \Theta, \hat \tau)$ divided by $\sigma^2$.
Let $v_{ij}$ denote the $(i,j)$ th element of $V$.
The standard $1-\alpha$ confidence interval for $\theta$ is
$I = \big [ \hat \Theta - t_{n-p,1-\frac{\alpha}{2}} \sqrt{v_{11}} \hat
\sigma, \quad \hat \Theta + t_{n-p,1-\frac{\alpha}{2}} \sqrt{v_{11}} \hat
\sigma \big ]$,
%
%
where the quantile $t_{m,a}$ is defined by $P(T \le t_{m,a}) = a$ for $T \sim t_m$ and
$\hat \sigma^2 = (Y - X \hat \beta)^T (Y - X \hat \beta)/(n-p)$.

%
%

The naive $1-\alpha$ confidence interval for $\theta$ is obtained as follows.
The usual test statistic for testing the null hypothesis that $\tau = 0$ against the alternative
hypothesis that $\tau \ne 0$ is $\hat \tau/(\hat \sigma \sqrt{v_{22}})$.
Suppose that, for some given positive number $q$, we fix $\tau$ at 0 if
$|\hat \tau|/(\hat \sigma \sqrt{v_{22}}) \le q$;
otherwise we allow $\tau$ to vary freely.
We use the notation $[a \pm b]$ for the interval $[a-b, a+b]$ ($b > 0$).
Also define $\rho = v_{12}/\sqrt{v_{11} v_{22}}$. Note that $\rho$ is the correlation
between $\hat \Theta$ and $\hat \tau$ and so it satisfies
$-1 \le \rho \le 1$.
The naive $1-\alpha$ confidence interval is as follows (Kabaila and Giri (2009a)).
If $|\hat \tau|/(\hat \sigma \sqrt{v_{22}}) > q$ then this confidence interval is
$\big [ \hat \Theta - t_{n-p,1-\frac{\alpha}{2}} \sqrt{v_{11}} \hat
\sigma, \quad \hat \Theta + t_{n-p,1-\frac{\alpha}{2}} \sqrt{v_{11}} \hat
\sigma \big ]$.
%
%
If, on the other hand, $|\hat \tau|/(\hat \sigma \sqrt{v_{22}}) \le q$ then this
confidence interval is
\begin{equation*}
\left [ \hat \Theta - \frac{v_{12}}{v_{22}} \hat \tau \, \pm \, t_{n-p+1,1-\frac{\alpha}{2}}
\sqrt{\frac{(n-p) \hat \sigma^2 + (\hat \tau^2/ v_{22})}{n-p+1}}
\sqrt{v_{11} - \frac{v_{12}^2}{v_{22}}}
\right ].
\end{equation*}
This confidence interval can be expressed in the new form
\begin{equation*}
\bigg [ \hat \Theta -
\sqrt{v_{11}} \hat \sigma \, b\bigg(\frac{\hat{\tau}}{\hat \sigma \sqrt{v_{22}}}\bigg)
\, \pm \, \sqrt{v_{11}} \hat \sigma \, s\bigg(\frac{|\hat{\tau}|}{\hat \sigma \sqrt{v_{22}}}\bigg)
\bigg ]
\end{equation*}
where
\begin{align*}
&b(x) =  \begin{cases}
        0 &\ \ \ \ \ \text{for } \ \ \ |x| > q\\
        \rho x &\ \ \ \ \ \text{for} \ \ \ \ |x| \le q.
    \end{cases} \\
&s(x) =  \begin{cases}
        t_{n-p,1-\frac{\alpha}{2}}  &\text{for } \ \ \ x > q\\
        t_{n-p+1,1-\frac{\alpha}{2}} \sqrt{1 -\rho^2} \sqrt{\frac{n-p + x^2}{n-p+1}}&\text{for} \ \ \ \ 0 < x \le q.
    \end{cases}
\end{align*}

In Section 4 we will consider the example of a $2 \times 2$
factorial experiment with 20 replicates. Here $p=4$. The parameter of interest
$\theta$ is the {\it simple} effect (expected response when factor A is high and factor B
is low) $-$ (expected response when factor A is low and factor B
is low).
We identify $\tau$ with the two-factor interaction, so that
$\rho = -1/\sqrt{2} = -0.7071068$.
Suppose that we have uncertain prior information that the two-factor interaction
is zero. Also suppose that we carry out a
preliminary test of the null hypothesis that the two-factor interaction is zero
against the alternative hypothesis that this interaction is non-zero.
Let the level of significance of this test be 0.05, so that $q=1.991673$.
Figure 1 is a plot of the functions $b$ and $s$ for the
resulting naive 0.95 confidence interval for $\theta$.
This confidence interval is completely inadequate, as its minimum coverage probability
is 0.7306. It also has the unpleasant feature that its endpoints are discontinuous functions
of the data.

\newpage
%

\begin{figure}[h]
\label{Figure1} \centering
\includegraphics[scale=0.75]{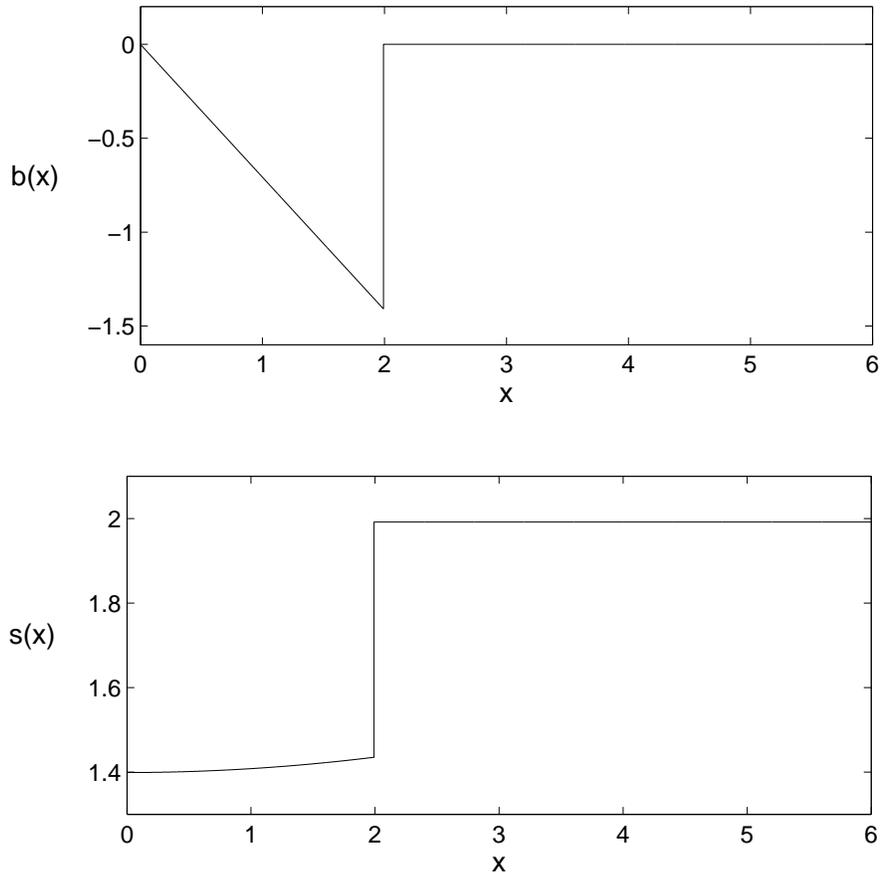}
       \caption{ Plots of the functions $b$ and $s$ for the
naive 0.95 confidence interval for the {\it simple} effect $\theta$ in the context of the
$2 \times 2$ factorial experiment with 20 replicates. This confidence interval is based on a preliminary
test of the null hypothesis that the two-factor interaction is zero against the alternative hypothesis
that this interaction is non-zero, with level of significance 0.05.}
\end{figure}


\newpage

\noindent {\large {\bf 3. New confidence interval utilizing prior information}}

\medskip

In this section we describe a broad class of confidence intervals for $\theta$. These confidence intervals
are required to have endpoints that are smooth function of the data. They are also required to coincide
with the standard $1-\alpha$ confidence intervals when the data strongly contradict the prior information.
We provide computationally
convenient expressions for the coverage probability and the scaled expected length for confidence intervals
from this class. These computationally convenient expressions were first described by Kabaila and Giri (2007a,b).
We then describe a weight function for the difference {\big (}(scaled expected length of the confidence interval) $-$
(scaled expected length of the standard $1-\alpha$ confidence interval){\big )}. This weight function gives the largest weight
to this difference when $\tau=0$ i.e. when the prior information is correct.
We find an interval that is optimal in the sense that
it minimizes the weighted average of this
difference subject to the constraint that it has minimum coverage probability $1-\alpha$.
Our choice of the weight function ensures that this interval utilizes the
prior information.

We introduce a confidence interval for $\theta$ that is similar in form to the naive $1-\alpha$ confidence interval,
described in the previous section, but with a great ``loosening up'' of the forms that the functions $b$ and $s$ can take.
Define the following confidence interval
for $\theta$
\begin{align}
\label{J(b,s)}
J(b,  s) = \bigg [ \hat \Theta -
\sqrt{v_{11}} \hat \sigma \, b\bigg(\frac{\hat{\tau}}{\hat \sigma \sqrt{v_{22}}}\bigg) \, \pm \,
\sqrt{v_{11}} \hat \sigma \, s\bigg(\frac{|\hat{\tau}|}{\hat \sigma \sqrt{v_{22}}}\bigg)
\bigg ]
\end{align}
where the functions $b$ and $s$ are required to satisfy the following restriction.
\smallskip

\noindent {\underbar{Restriction 1}}
\newline $b: \mathbb{R} \rightarrow \mathbb{R}$  is constrained to be
an odd function and $s: [0, \infty)
\rightarrow [0, \infty)$.

\smallskip
\noindent The motivation for restricting attention to this form of
interval is provided by the new invariance arguments presented in Appendix A.
We also require that the functions $b$ and $s$
satisfy the following restriction.
\smallskip

\noindent {\underbar{Restriction 2}} \newline
$b$ and $s$ are continuous functions.

\smallskip
\noindent This implies that the endpoints of the confidence interval $J(b,s)$ are continuous functions
of the data. Finally, we require the confidence interval $J(b,s)$ to coincide with
the standard $1-\alpha$ confidence interval $I$ when the data strongly contradict
the prior information. The statistic $|\hat \tau|/(\hat \sigma \sqrt{v_{22}})$ provides some indication
of how far away $\tau / (\sigma \sqrt{v_{22}})$ is from 0.
We therefore require that
the functions $b$ and $s$
satisfy the following restriction.
\smallskip

\noindent {\underbar{Restriction 3}} \newline
$b(x)=0$ for all $|x| \ge d$ and $s(x)=t_{n-p,1-\frac{\alpha}{2}}$ for all $x \ge d$
where $d$ is a (sufficiently large) specified positive number.

\medskip

Define $\gamma = \tau/(\sigma \sqrt{v_{22}})$,
$G = (\hat \Theta - \theta)/(\sigma \sqrt{v_{11}})$
and $H = \hat \tau/(\sigma \sqrt{v_{22}})$. Note that
\begin{equation}
\label{model_G_H}
\left[\begin{matrix} G\\ H \end{matrix}
\right] \sim N \left ( \left[\begin{matrix} 0 \\ \gamma \end{matrix}
\right], \left[\begin{matrix} 1 \quad \rho\\ \rho \quad 1 \end{matrix}
\right] \right ).
\end{equation}
where, as defined in Section 2, $\rho = v_{12}/\sqrt{v_{11} v_{22}}$.
Also define $W = \hat \sigma/\sigma$.
Note that $(G,H)$ and $W$ are independent random
vectors. Also, $W$ has the same distribution as $\sqrt{Q/(n-p)}$ where $Q
\sim \chi^2_{n-p}$.
Let $f_W$ denote the probability density function of $W$.

It is straightforward to show that
the coverage probability $P\big( \theta \in J(b, s) \big)$
is equal to
$P \big ( \ell(H,W) \le G \le u(H,W) \big )$,
%
%
where the functions $\ell(\cdot, \cdot) : \mathbb{R} \times [0, \infty) \rightarrow \mathbb{R}$
and $u(\cdot, \cdot) : \mathbb{R} \times [0, \infty) \rightarrow \mathbb{R}$  are defined by
$\ell(h,w) = b(h/w)\, w - s(h/w) \, w$  and
$u(h,w) =  b(h/w)\, w + s(h/w) \, w$.
%
%
For given $b$, $s$ and $\rho$, the coverage probability of $J(b,s)$ is a function of $\gamma$.
We denote this coverage probability by $c(\gamma;b,s, \rho)$.

Part of our evaluation of the confidence interval $J(b,s)$ consists of comparing it with the
standard $1-\alpha$ confidence interval
$I$ using the criterion
\begin{equation}
\label{criterion_initial}
\frac{\text{expected length of $
J(b, s)$}}
{\text{expected length of $I$}}.
\end{equation}
We call this the scaled expected length of $J(b, s)$. This is equal to
\begin{equation*}
\frac{E \left ( s \bigg( \displaystyle{\frac{|H|}{W}}\bigg) W \right )}
{t_{n-p,1-\frac{\alpha}{2}} E(W)}.
\end{equation*}
This is a function of $\gamma$ for given $s$. We denote this function by $e(\gamma;s)$.
Clearly, for given $s$, $e(\gamma;s)$ is an even function of $\gamma$.

Our aim is to find functions $b$ and $s$ that satisfy Restrictions 1--3 and such that (a) the minimum of $c(\gamma;b,s, \rho)$ over $\gamma$ is
$1-\alpha$ and (b)
\begin{equation}
\label{criterion}
\int_{-\infty}^{\infty} (e(\gamma;s) - 1) \, d \nu(\gamma)
\end{equation}
is minimized, where the weight function $\nu$ has been chosen to be
\begin{equation}
\label{mixed_wt_fn}
\nu(x) = \lambda x + {\cal H}(x) \ \text{ for all } \ x \in \mathbb{R},
\end{equation}
where $\lambda$ is a specified nonnegative number and
${\cal H}$ is the unit step function defined by
${\cal H}(x) = 0$ for $x<0$ and ${\cal H}(x) = 1$ for
$x \ge 0$.
%
%
The larger the value of $\lambda$, the smaller the relative weight given to
minimizing $e(\gamma;s)$ for $\gamma=0$, as opposed to
minimizing $e(\gamma;s)$ for other values of $\gamma$.
Similarly to Farchione and Kabaila (2008), who consider a much simpler model,
we expect the weight function \eqref{mixed_wt_fn} to lead to a $1-\alpha$
confidence interval for $\theta$ that has expected length that (a) is relatively small
when $\tau=0$ and (b) has maximum value that is not too large.

The following theorem provides new computationally convenient expressions
for the coverage probability and scaled expected length of $J(b, s)$.

\medskip

\noindent {\bf Theorem 1.}

\noindent (a) Define the functions
$k^{\dag}(h,w, \gamma, \rho) = \Psi \big( -t_{n-p,1-\frac{\alpha}{2}} w, t_{n-p,1-\frac{\alpha}{2}} w;
 \rho(h-\gamma), 1-\rho^2 \big )$ and
$k(h,w,\gamma, \rho) = \Psi \big(\ell(h,w), u(h,w); \rho(h-\gamma),1-\rho^2 \big )$,
%
%
%
where $\Psi(x, y; \mu, v) = P(x \le Z \le y)$ for $Z \sim N(\mu,v)$.
The coverage probability of $J(b, s)$ is denoted by $c(\gamma;b,s, \rho)$ and is equal to
\begin{equation}
\label{cov_prob}
(1-\alpha) +
\int_0^{\infty} \int_{-d}^{d} \big( k(wx,w, \gamma, \rho) - k^{\dag}(wx,w, \gamma, \rho) \big)
\, \phi(wx-\gamma)\,  dx \, w \, f_W(w) \, dw
\end{equation}
where $\phi$ denotes the $N(0,1)$ probability density function.
For given $b$, $s$ and $\rho$, $c(\gamma;b,s, \rho)$ is an even function of $\gamma$.

\smallskip

\noindent (b) The scaled expected length of $J(b, s)$
is
\begin{equation}
\label{comp_conv_e}
e(\gamma;s) = 1 +
 \frac{1} {t_{n-p, 1 - \frac{\alpha}{2}} \, E(W)}
 \int^{\infty}_0 \int^{d}_{- d} \left (s(|x|) - t_{n-p,1-\frac{\alpha}{2}} \right )
\phi(w x -\gamma) \, dx \,  w^2 \,  f_W(w)  \, dw.
\end{equation}

\medskip

Substituting \eqref{comp_conv_e} into \eqref{criterion}, we obtain that \eqref{criterion} is equal to
\begin{align}
\label{criterion_final}
 &\frac{1} {t_{n-p, 1 - \frac{\alpha}{2}} \, E(W)} \int_{-\infty}^{\infty}
 \int^{\infty}_0 \int^{d}_{- d} \left (s(|x|) - t_{n-p,1-\frac{\alpha}{2}} \right )
\phi(w x -\gamma) \, dx \,  w^2 \,  f_W(w)  \, dw \, d \nu (\gamma) \notag \\
&=\frac{1} {t_{n-p, 1 - \frac{\alpha}{2}} \, E(W)}
 \int^{\infty}_0 \int^{d}_{- d} \left (s(|x|) - t_{n-p,1-\frac{\alpha}{2}} \right )
\int_{-\infty}^{\infty} \phi(w x -\gamma) \, d \nu (\gamma)\, dx \,  w^2 \,  f_W(w)  \, dw \notag \\
&=\frac{2} {t_{n-p, 1 - \frac{\alpha}{2}} \, E(W)}
 \int^{\infty}_0 \int^{d}_{0} \left (s(x) - t_{n-p,1-\frac{\alpha}{2}} \right )
(\lambda + \phi(w x)) \, dx \,  w^2 \,  f_W(w)  \, dw
\end{align}

For computational feasibility, we specify the following parametric forms for the
functions $b$ and $s$.
We require $b$ to be a continuous function and so it is necessary that
$b(0)=0$.
Suppose that $x_1, \ldots, x_q$
satisfy $0 = x_1 < x_2 < \cdots < x_q = d$.
Obviously, $b(x_1)=0$, $b(x_q)=0$ and $s(x_q)=t_{n-p,1-\frac{\alpha}{2}}$.
The function $b$ is fully specified by the vector $\big (b(x_2), \ldots, b(x_{q-1}) \big)$
as follows. Because $b$ is assumed to be an odd function, we know that
$b(-x_i) = -b(x_i)$ for $i=2,\ldots,q$.
We specify the value of $b(x)$ for any $x \in [-d,d]$
by cubic spline interpolation for these given function values, subject to the constraint
that $b^{\prime}(-d)=0$ and $b^{\prime}(d)=0$.
We fully specify the function $s$ by the vector $\big (s(x_1), \ldots, s(x_{q-1}) \big)$
as follows.
The value of $s(x)$ for any $x \in [0,d]$ is specified
by cubic spline interpolation for these given function values (without any endpoint conditions
on the first derivative of $s$). We call $x_1, x_2, \ldots x_q$ the knots.

To conclude, the new $1-\alpha$ confidence interval for $\theta$ that utilizes the prior information
that $\tau=0$ is obtained as follows. For a judiciously-chosen set of values
of $d$, $\lambda$ and knots $x_i$, we carry out the following computational procedure.

\smallskip

\noindent {\underbar{Computational Procedure}} \newline
Compute the functions $b$ and $s$, satisfying Restrictions 1--3 and taking the
parametric forms described above, such that (a) the minimum over $\gamma \ge 0$ of
\eqref{cov_prob} is $1-\alpha$ and (b) the criterion \eqref{criterion_final} is minimized.
Plot $e^2(\gamma;s)$, the square of the scaled expected length,
as a function of $\gamma \ge 0$.

\smallskip

\noindent Based on these plots and the strength of our prior information that $\tau=0$,
we choose appropriate values of $d$, $\lambda$ and knots $x_i$.
The confidence interval corresponding to this choice is the new
$1-\alpha$ confidence interval for $\theta$.

\medskip

\noindent {\underbar{\sl Remark 3.1}} \ Suppose that $\lambda > 0$ is fixed.
Also suppose that we apply the Computational Procedure without any parametric restrictions
of the form described above.
The structure of the criterion \eqref{criterion}
when $\nu$ is given by \eqref{mixed_wt_fn} make
it highly plausible that the resulting
 $1-\alpha$ confidence interval for $\theta$ will have
a scaled expected length $e(\gamma;s)$
that converges uniformly in $\gamma$ to some limiting function as $d \rightarrow \infty$.
It is also highly plausible that this limiting function can be found to a very
good approximation by applying this Computational Procedure for
$d$ sufficiently large and knots $x_i$ sufficiently closely spaced.

\bigskip

\noindent {\large { \bf {4. Application to the analysis of data from a $\mathbf{2 \times 2}$ factorial \newline
\phantom{123}experiment}}}

\medskip

In this section we consider a $2 \times 2$ factorial experiment with
20 replicates and parameter of interest $\theta$
the {\it simple} effect (expected response when factor A is high and factor B is low) $-$
(expected response when factor A is low and factor B is low).
We suppose that we have uncertain prior information that the two-factor
interaction is zero.
We use this example to illustrate the properties of the new $1-\alpha$
confidence interval for $\theta$ that utilizes this prior information, when
$1-\alpha=0.95$. All of the computations presented in this paper were performed with programs
written in MATLAB, using the Optimization and Statistics toolboxes.

Let $x_1$ take the values $-1$ and 1 when the factor A takes the values
low and high respectively. Also let $x_2$ take the values $-1$ and 1 when the factor B takes the values
low and high respectively. In other words, $x_1$ and $x_2$ are the coded values of the
factors A and B respectively. The model for this experiment is
\begin{equation}
\label{regression}
Y = \beta_0 + \beta_1 x_1 + \beta_2 x_2 + \beta_{12} x_1 x_2 + \varepsilon
\end{equation}
where $Y$ is the response, $\beta_0$, $\beta_1$, $\beta_2$ and $\beta_{12}$ are unknown parameters and the
$\varepsilon$ for different response measurements are independent and identically $N(0, \sigma^2)$ distributed.
Thus $\theta = 2(\beta_1 - \beta_{12})$. Let $\hat \beta_1$ and $\hat \beta_{12}$ denote the least squares
estimators of $\beta_1$ and $\beta_{12}$ respectively. The least squares estimator of $\theta$ is
$\hat \Theta = 2(\hat \beta_1 - \hat \beta_{12})$. Our uncertain prior information is that $\beta_{12}=0$. Note that
\begin{equation*}
\left[\begin{matrix} \hat \Theta\\ \hat \beta_{12} \end{matrix}
\right] \sim N \left ( \left[\begin{matrix} \theta \\
\beta_{12} \end{matrix}
\right], \frac{\sigma^2}{80} \left[\begin{matrix} \phantom{1}8 \quad -2 \\ -2 \quad \ \ 1 \end{matrix}
\right] \right ).
\end{equation*}
Hence $\rho = -1/\sqrt{2}$.

We followed the Computational Procedure, described at the end of
the previous section, with $d=6$, $\lambda=0.2$ and evenly-spaced
knots $x_i$ at $0, 1, 2, \ldots, 6$. The resulting functions $b$
and $s$, which specify the new 0.95 confidence interval for
$\theta$, are plotted in Figure 2. The performance of this
confidence interval is shown in Figure 3. This confidence interval
has coverage probability 0.95 throughout the parameter space. When
the prior information is correct (i.e. $\gamma=0$), we gain since
$e^2(0;s) = 0.8683$. The maximum value of $e^2(\gamma;s)$ is
1.1070. This confidence interval coincides with the standard
$1-\alpha$ confidence interval for $\theta$ when the data strongly
contradicts the prior information, so that $e^2(\gamma;s)$
approaches 1 as $\gamma \rightarrow \infty$. It is interesting to
note the broad qualitative similarities between the functions
plotted in Figures 1 and 2.

These values of $d=6$, $\lambda=0.2$ and knots $x_i$ were obtained
after a search that we summarize as follows. Consider $d=6$,
evenly-spaced knots $x_i$ at $0, 1, 2, \ldots, 6$ and $\lambda =
0.05$, 0.2 , 0.5 and 1. The Computational Procedure was applied
for each of these values. As expected from the form of the weight
function, for each of these values of $\lambda$, $e^2(\gamma;s)$
is minimized at $\gamma=0$. For a given value of $\lambda$, define
the `expected gain' to be $\big(1 - e^2(0;s) \big)$ and the
`maximum potential loss' to be $\big (\max_{\gamma} e^2(\gamma;s)
- 1 \big)$. As shown in Table 1, as $\lambda$ increases (a) the
expected gain decreases and (b) the ratio (expected gain)/(maximum
potential loss) increases. By choosing $\lambda=0.2$ we have both
a reasonably large expected gain and a reasonably large value of
the ratio (expected gain)/(maximum potential loss).

\medskip

\begin{table}[h]
\renewcommand{\arraystretch}{1.3}
\hfil
\begin{tabular}{|c|c|c|c|c|c|c|c|c|}
  \hline
  $\lambda$ & 0.05 & 0.2 & 0.5& 1 \\
  \hline
  expected gain & 0.196 & 0.1317 & 0.0822& 0.043   \\
  \hline
 maximum potential loss & 0.2610 & 0.1070 & 0.0503 & 0.0248 \\
  \hline
  (expected gain)/(maximum potential loss) & 0.7509 & 1.2308 & 1.6341 & 1.7338 \\
  \hline
\end{tabular}
\hfil \caption{Performance of the new 0.95 confidence interval for
$d=6$ and knots $x_i$ at $0, 1, \ldots, 6$ when we vary over
$\lambda \in \{0.05, 0.2, 0.5, 1\}$.} \label{size_table}
\end{table}
\FloatBarrier

\medskip

Now consider $\lambda=0.2$ and evenly-spaced knots $x_i$ at $0, 1, 2, \ldots, d$ where $d = 4$, 6, 8 and 10.
The Computational Procedure was applied for each
of these values. There was a marked improvement in performance of the resulting 0.95 confidence
interval when $d$ was increased from 4 to 6. However, the improvement in performance
of the resulting 0.95 confidence was negligible when $d$ was increased from 6 to 8 and from 6 to 10.
This suggests that increasing $d$ beyond 6 will lead to a negligible improvement in performance of
the confidence interval.

Finally, consider $d=6$, $\lambda = 0.2$ and two sets of evenly-spaced
knots $x_i$ at $0, 0.6, 1.2, 1.8, \ldots, 6$
and $0, 0.5, 1, 1.5, \ldots, 6$.
The Computational Procedure was applied to both of these sets of knots.
The improvements in performance of the resulting 0.95 confidence interval
(compared to the performance for $d=6$, $\lambda = 0.2$ and evenly-spaced knots $x_i$ at $0, 1, 2, \ldots, 6$)
were practically negligible. This suggests that there will be a practically negligible improvement in performance
if the spacing between the evenly-spaced knots is reduced to less than 1.

\begin{figure}[h]
\label{Figure1} \centering
\includegraphics[scale=0.75]{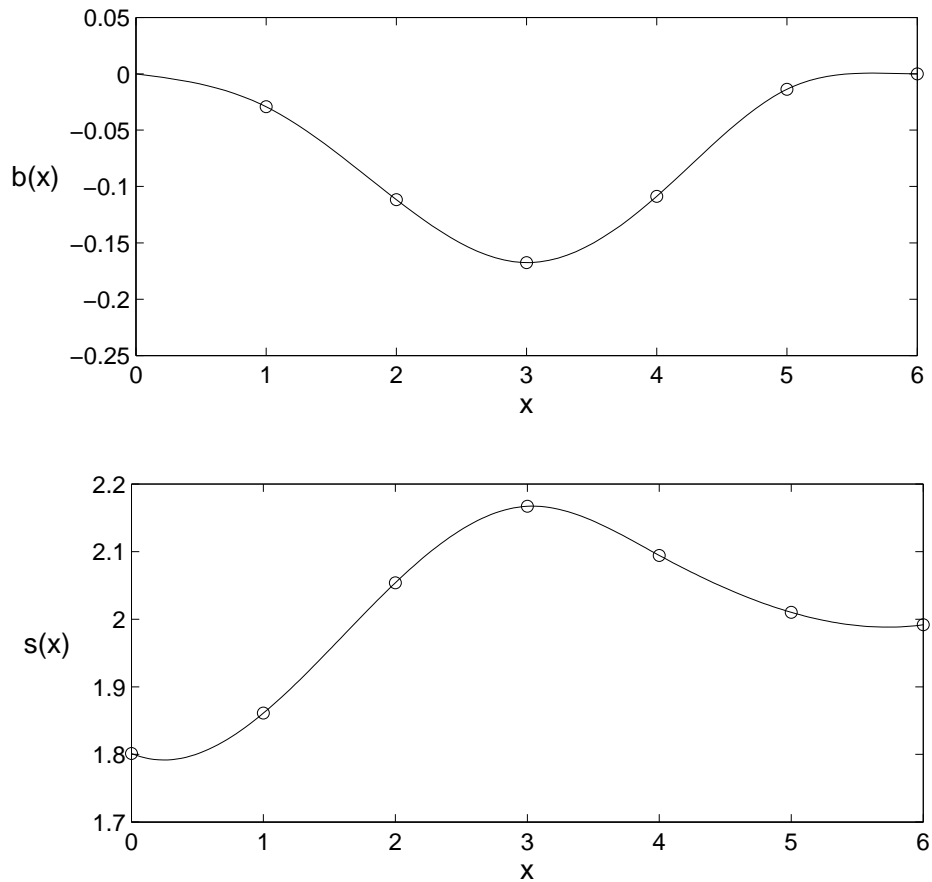}
       \caption{ Plots of the functions $b$ and $s$ for the
       new $1-\alpha$ confidence interval in the context of a
       $2 \times 2$ factorial experiment
       with 20 replicates, parameter of interest the {\it simple} effect
       $\theta = 2(\beta_1 - \beta_{12})$ and $1-\alpha=0.95$.
       These functions were obtained using $d=6$, $\lambda=0.2$ and the knots $x_i$ at $0, 1, 2, \ldots, 6$.}
\end{figure}
\newpage
\FloatBarrier
\begin{figure}[h]
\label{Figure1} \centering
\includegraphics[scale=0.75]{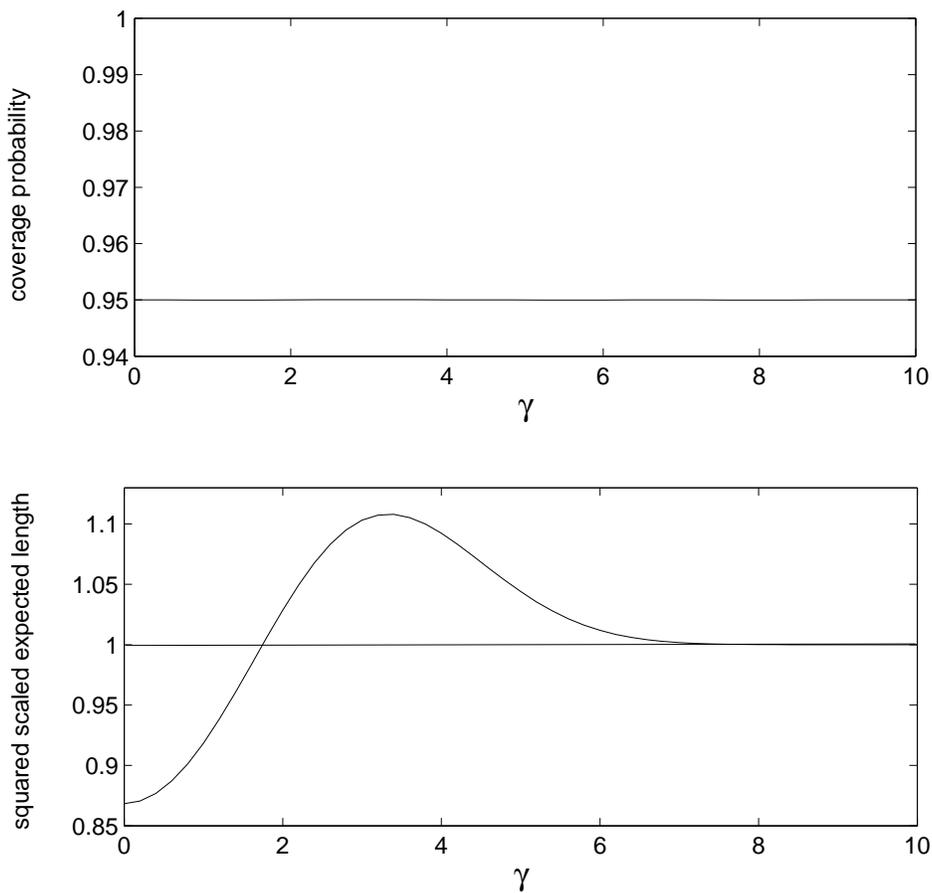}
       \caption{ Plots of the coverage probability and $e^2(\gamma;s)$, the squared scaled expected length,
{\Big(}as functions of $\gamma = \beta_{12}/\sqrt{\text{var}(\hat \beta_{12})}${\Big )} of the new 0.95
confidence interval for the {\it simple} effect $\theta = 2(\beta_1 - \beta_{12})$ for the $2 \times 2$ factorial experiment
with 20 replicates.  These functions were obtained using $d=6$, $\lambda=0.2$ and the knots $x_i$ at $0, 1, 2, \ldots, 6$.}
\end{figure}
\FloatBarrier


\newpage

\noindent {\large{\bf {5. Discussion}}}

\medskip

\noindent {\underbar{\sl Discussion 5.1}} \ Our motivation for the weight function \eqref{mixed_wt_fn}
is as follows. Suppose that the only restriction on the functions $b: \mathbb{R} \rightarrow \mathbb{R}$
and $s: [0, \infty) \rightarrow [0, \infty)$ is that $b$ is an odd function. Consider the weight function
$\nu = {\cal H}$, which corresponds to all of the weight being placed at $\tau=0$. The minimization of
\eqref{criterion}, subject to $P\big( \theta \in J(b, s) \big) \ge 1-\alpha$ for all $\gamma$, leads to
a $1-\alpha$ confidence interval for $\theta$ with the following properties. This interval has the smallest
expected length when $\tau=0$ (i.e. when the prior information is correct) of any $1-\alpha$ confidence interval
for $\theta$. However, this confidence interval has the weakness that its expected length
approaches infinity as $|\gamma| \rightarrow \infty$ (Tuck, 2006). Now consider the weight function
$\nu = x$, which corresponds to a uniform weight over $\mathbb{R}$. The minimization of
\eqref{criterion}, subject to $P\big( \theta \in J(b, s) \big) \ge 1-\alpha$ for all $\gamma$, leads to
the standard $1-\alpha$ confidence interval $I$. Finally, consider the weight function
\eqref{mixed_wt_fn}, which is a mixture of the weight functions ${\cal H}$ and $x$, for fixed $\lambda > 0$.
This weight function puts a large amount of weight at $\tau = 0$, consistent with our desire that
the confidence interval has relatively small expected length when the prior information is correct.
Also, the $x$ component of this weight function leads to a confidence interval whose expected length
has a maximum value that is finite. In addition, the structure of the criterion \eqref{criterion}
when $\nu$ is given by \eqref{mixed_wt_fn} makes it highly plausible that the $1-\alpha$ confidence interval
resulting from the minimization of \eqref{criterion},
 subject to $P\big( \theta \in J(b, s) \big) \ge 1-\alpha$ for all $\gamma$,
 will have the desirable feature that it approaches the standard $1-\alpha$ confidence interval $I$
 as the data increasingly contradict the prior information.
 Fortuitously, this property leads to the computational advantage described in Remark 3.1.

\medskip

\noindent {\underbar{\sl Discussion 5.2}} \ The new $1-\alpha$ confidence interval is computed to satisfy
the constraint that its minimum coverage probability is $1-\alpha$.
For the example described in Section 4,
it is remarkable that the new $1-\alpha$ confidence interval has coverage probability {\sl equal} to $1-\alpha$
throughout the parameter space. The new $1-\alpha$ confidence interval has been computed for a wide
range of values of $1-\alpha$, $\lambda$, $\rho$, $n-p$ (including the limiting case $n-p \rightarrow \infty$),
$d$ and knots $x_i$. In each case,
the new $1-\alpha$ confidence interval has coverage probability equal to $1-\alpha$
throughout the parameter space. This provides strong empirical evidence that the
new $1-\alpha$ confidence interval has the attractive property that its coverage probability is equal to $1-\alpha$
throughout the parameter space.

\medskip

\noindent {\underbar{\sl Discussion 5.3}} \ The new $1-\alpha$ confidence interval has been computed for a wide
range of values of $1-\alpha$, $\lambda$, $\rho$, $n-p$ (including the limiting case $n-p \rightarrow \infty$),
$d$ and knots $x_i$.
For each of these values of $1-\alpha$, $\lambda$, $d$ and knots $x_i$, $e^2(0;s)$ (which is the
minimum value of $e^2(\gamma;s)$) decreases when $|\rho|$ increases and/or $(n-p)$ decreases.

\medskip

\noindent {\underbar{\sl Discussion 5.4}} \ Consider the particular case that $\rho = 0$. In this case,
we expect that any improvement
in performance of the new $1-\alpha$ confidence interval
over the standard $1-\alpha$ confidence interval $I$ can only be due to improved estimation of
the parameter $\sigma$. Computations
show that the new $1-\alpha$ confidence interval performs well (in terms of utilizing the
uncertain prior information)
for small $n-p$, when $\lambda$ is chosen appropriately.
However, the new $1-\alpha$ confidence interval
approaches the standard $1-\alpha$ confidence interval $I$ as $n-p \rightarrow \infty$.

\medskip

\noindent {\underbar{\sl Discussion 5.5}} \ We briefly compare our frequentist approach with a
Bayesian approach to the problem stated in the paper. A full discussion will be presented in a
separate paper. For simplicity, suppose that $\sigma^2$
is known and that
\begin{equation*}
\left[\begin{matrix} \hat \Theta\\ \hat \tau \end{matrix}
\right] \sim N \left ( \left[\begin{matrix} \theta \\ \tau \end{matrix}
\right], \left[\begin{matrix} 1 \quad \rho\\ \rho \quad 1 \end{matrix}
\right] \right ).
\end{equation*}
For the Bayesian approach, suppose that we choose independent prior pdf's for $\Theta$ and
$\tau$. Also suppose that for this approach (a) $\Theta$ has an uniform improper prior pdf and (b) $\tau$ has the
prior pdf $\xi \delta(\tau) + (1-\xi)$ where $\delta$ denotes the delta function and
$\xi$ is a fixed number satisfying $0 \le \xi \le 1$. Contrasting features of the
new frequentist $1-\alpha$ confidence interval for $\theta$ described in the present paper
and the Bayesian $1-\alpha$ highest probability density (HPD) regions for $\Theta$ include the following:

\smallskip

\noindent (a) Suppose that the only restriction on the functions $b: \mathbb{R} \rightarrow \mathbb{R}$
and $s: [0, \infty) \rightarrow [0, \infty)$ is that $b$ is an odd function. Consider the weight function
$\nu = {\cal H}$, which corresponds to all of the weight being placed at $\tau=0$. The minimization of
\eqref{criterion}, subject to $P\big( \theta \in J(b, s) \big) \ge 1-\alpha$ for all $\gamma$, leads to
a $1-\alpha$ confidence interval with the smallest
expected length when $\tau=0$ of any $1-\alpha$ confidence interval
for $\theta$. There is {\sl no Bayesian analogue} of this confidence interval. If we choose $\xi = 1$
then the Bayesian $1-\alpha$ HPD region for $\Theta$ is equal to the usual $1-\alpha$ confidence interval for
$\theta$ based on the assumption that $\tau = 0$. This confidence interval has coverage probability
with infimum 0.

\smallskip

\noindent (b) By the appropriate choices of $1-\alpha$, $\xi$, $\rho$, $\sigma$ and $\hat \tau$, one
can find Bayesian $1-\alpha$ HPD regions for $\Theta$ that consist of the union of two disjoint
intervals. By contrast, the methodology of the present paper always produces a confidence interval.

\smallskip

\noindent (c) By the appropriate choices of $1-\alpha$, $\xi$ where $\xi < 1$, $\rho$ and $\sigma$, one
can find Bayesian $1-\alpha$ HPD regions for $\Theta$ that have frequentist minimum coverage probabilities
far below $1-\alpha$.

\medskip

\noindent {\underbar{\sl Discussion 5.6}} \ We briefly discuss the computation of the new confidence interval.
A full discussion is provided by Giri (2008) and will be presented in a separate paper.
Our first step has been to truncate the integrals
with respect to $w$ in \eqref{cov_prob}, \eqref{comp_conv_e} and \eqref{criterion_final} and to find upper
bounds on the truncation errors. The computational implementation of the constraints that
$c(\gamma;b,s, \rho) \ge 1 - \alpha$ for all $\gamma \ge 0$ is as follows. Restriction 3 implies that, for
any reasonable choice of the functions $b$ and $s$, $c(\gamma;b,s, \rho) \rightarrow 1-\alpha$
as $\gamma \rightarrow \infty$. The constraints implemented in the computer programs are that
$c(\gamma;b,s, \rho) \ge 1 - \alpha$ for each $\gamma \in \{0, \Delta, 2 \Delta, \ldots, M \Delta \}$ where
$\Delta$ is sufficiently small and $M$ is sufficiently large.

\medskip

\noindent {\underbar{\sl Discussion 5.7}} \ The new $1-\alpha$ confidence interval for $\theta$ is founded
on the assumption that the random errors $\varepsilon_i$ are independent and identically $N(0, \sigma^2)$
distributed. This confidence interval is based on the least squares estimator $\hat \Theta$ of $\theta$
and the estimator $\hat \sigma$ of $\sigma$. Consequently, it will display the same kind of lack of robustness
to non-normality of the random errors as the standard $1-\alpha$ confidence interval $I$.

\medskip

\noindent {\underbar{\sl Discussion 5.8}} \ We illustrate our method with the following real data set.
We extract a $2 \times 2$ factorial data set from the $2^3$ factorial data set described in Table 7.5
of Box et al (1963) as follows. Define $x_1 = -1$ and $x_1 = 1$ for ``Time of addition of HNO$_3$'' equal
to 2 hours and 7 hours, respectively.  Also define $x_2 = -1$ and $x_2 = 1$ for ``heel absent'' and
``heel present'', respectively. The observed responses are the following: $y=87.2$ for $(x_1,x_2) = (-1,-1)$,
$y=88.4$ for $(x_1,x_2) = (1,-1)$, $y=86.7$ for $(x_1,x_2) = (-1,1)$ and $y=89.2$ for $(x_1,x_2) = (1,1)$.
We use the model \eqref{regression}. The discussion on p.265 of Box et al (1963) implies that there is
uncertain prior information that $\beta_{12} = 0$. The discussion on p.266 of Box et al (1963) implies that there is
an estimator $\hat \sigma^2$ of $\sigma^2$, obtained from other related experiments, with the property
that $\hat \sigma^2/\sigma^2 \sim Q/m$ where $Q \sim \chi^2_m$ and $m$ is effectively infinite.
The observed value of $\hat \sigma$ is 0.8. As in Section 4, define the parameter of interest $\theta$
to be the simple effect (expected response when $x_1 = 1$ and $x_2 = -1$)
$-$ (expected response when $x_1 = -1$ and $x_2 = -1$), so that $\theta = 2(\beta_1 - \beta_{12})$.
Thus
\begin{equation*}
\left[\begin{matrix} \hat \Theta\\ \hat \beta_{12} \end{matrix}
\right] \sim N \left ( \left[\begin{matrix} \theta \\
\beta_{12} \end{matrix}
\right],\sigma^2 \left[\begin{matrix} \phantom{1}2 \quad -1/2 \\ -1/2 \quad \ \ 1/4 \end{matrix}
\right] \right ).
\end{equation*}
The standard 0.95 confidence interval for $\theta$ is $[-1.01745, 3.41745]$.
We have also computed the new 0.95 confidence interval for $\theta$
using $d=6$, $\lambda = 0.2$ and equally-spaced knots at $0, 6/8, \ldots, 6$. This confidence interval
is $[-0.81967, 3.26345]$,
which is substantially shorter than the standard 0.95 confidence interval.

\medskip

\noindent {\underbar{\sl Discussion 5.9}} \ Denote the the usual
$1-\alpha$ confidence interval for $\theta$, based on the assumption that $\tau = 0$,
by $K$. The naive $1-\alpha$ confidence interval described in Section 2
may be viewed as being obtained via a monotone discontinuous transition, based on the value of the test statistic
$|\hat \tau|/(\hat \sigma \sqrt{v_{22}})$, from the standard $1-\alpha$ confidence interval $I$ to $K$.
What are the properties of the confidence interval that results from replacing this monotone discontinuous transition
by a monotone continuous transition?

For simplicity, consider the case that $n-p$ is large.
Define the quantile $z_a$ by $P(Z \le z_a) = a$ for $Z \sim N(0,1)$.
In this case,
$I = \big [ \hat \Theta - z_{1-\frac{\alpha}{2}} \sqrt{v_{11}} \hat
\sigma, \quad \hat \Theta + z_{1-\frac{\alpha}{2}} \sqrt{v_{11}} \hat
\sigma \big ]$
%
%
and
$K = \big [ \hat \Theta - (\hat \tau/(\hat \sigma \sqrt{v_{22}})) \rho
\sqrt{v_{11}} \hat \sigma \pm
z_{1-\frac{\alpha}{2}} \sqrt{v_{11}} \hat \sigma \sqrt{1-\rho^2} \big ]$.
%
%
The naive $1-\alpha$ confidence interval for $\theta$ described in Section 2 may be
expressed in the following form
\begin{equation}
\label{mixture}
g \left(\frac{|\hat \tau|}{\hat \sigma \sqrt{v_{22}}} \right ) I +
\left ( 1 - g \left(\frac{|\hat \tau|}{\hat \sigma \sqrt{v_{22}}} \right ) \right) K
\end{equation}
where $g: [0, \infty) \rightarrow [0,1]$ is the step function defined by
$g(x) = 0$ for all $x \in [0,q]$ and $g(x) = 1$ for all $x > q$.

Now suppose that, instead, $g$ is a continuous increasing function satisfying $g(0) = 0$ and $g(x) \rightarrow 1$
as $x \rightarrow \infty$.
What are the properties of the confidence interval \eqref{mixture} in this case?
It is straightforward to show that \eqref{mixture} can be expressed in the form
\eqref{J(b,s)} with
$b(x) = (1 - g(|x|)) \rho x$ for all $x \in \mathbb{R}$ and
$s(x) = \left (g(x) \big(1 - \sqrt{1-\rho^2} \big) + \sqrt{1-\rho^2} \right) z_{1-\frac{\alpha}{2}}$
for all $x \ge 0$.
%
%
In other words, the confidence interval \eqref{mixture} is of the form
\eqref{J(b,s)}, but with very severe constraints on the functions $b$ and $s$.
In particular, $s(0) = \sqrt{1-\rho^2} z_{1-\frac{\alpha}{2}}$ and
$s(x)$ is a nondecreasing function that converges to $z_{1-\frac{\alpha}{2}}$ as
$x \rightarrow \infty$. The new $1-\alpha$ confidence interval described in
Section 3 has been computed for a wide range of values of $\rho > 0$ and in every single
case these very severe constraints are far from satisfied by $s$.
So, the confidence interval \eqref{mixture}
does not provide a shortcut to finding the new confidence interval described in Section 3.
Indeed, the strength of these constraints on the functions $b$ and $s$ implies that
any confidence interval of the form \eqref{mixture} will be far inferior to the
new confidence interval described in Section 3.
The results of Joshi (1969) show that the confidence interval $I$ is admissible,
with the consequence that the minimum coverage probability
of the confidence interval \eqref{mixture} must be less than $1-\alpha$.



\bigskip

\noindent {\large{\bf {Appendix A. Invariance arguments}}}

\medskip

 In this appendix we provide a motivation for considering a confidence interval
for $\theta$ of the form \eqref{J(b,s)}
where $b: \mathbb{R} \rightarrow \mathbb{R}$  is constrained to be
an odd function and $s: [0, \infty)
\rightarrow [0, \infty)$.
We provide this motivation through the invariance arguments
listed below.
Traditional invariance arguments (see e.g. Casella and Berger (2002, section 6.4) do not include considerations
of the available prior information. The novelty in the present appendix is that the invariance arguments need
to take proper account of the prior information.
Suppose that we have uncertain prior information that $\tau=0$.
Remember that the parameter of interest $\theta$ is defined to be $a^T \beta$.

Our first step is to reduce the data to $(\hat \Theta, \hat \tau, \hat \sigma)$. Note that $(\hat \Theta, \hat \tau)$ and
$\hat \sigma$ are independent random vectors with
\begin{equation*}
\left[\begin{matrix} \hat \Theta\\ \hat \tau \end{matrix}
\right] \sim N \left ( \left[\begin{matrix} \theta \\
\tau \end{matrix}
\right], \sigma^2 V \right ).
\end{equation*}
and $(n-p) \hat \sigma^2/\sigma^2 \sim \chi_{n-p}^2$.
Consider a confidence interval
\begin{equation}
\tag{A.1}
\label{A.1}
\big [ \ell(\hat \Theta, \hat \tau, \hat \sigma),
u(\hat \Theta, \hat \tau, \hat \sigma) \big ]
\end{equation}
for $\theta$ where $\ell: \mathbb{R} \times \mathbb{R} \times [0,\infty) \rightarrow \mathbb{R}$
and $u: \mathbb{R} \times \mathbb{R} \times [0,\infty) \rightarrow \mathbb{R}$.

\medskip

\noindent {\bf Invariance Argument 1}

\noindent The model for the reduced data may be re-expressed
\begin{equation*}
\left[\begin{matrix} \hat \Theta^{\dag}\\ \hat \tau^{\dag} \end{matrix}
\right] \sim N \left ( \left[\begin{matrix} \theta^{\dag} \\
\tau^{\dag} \end{matrix}
\right], (\sigma^{\dag})^2 V \right ).
\end{equation*}
where $\theta^{\dag} = \theta + c$, $\tau^{\dag} = \tau$, $\sigma^{\dag} = \sigma$,
$\hat \Theta^{\dag} = \hat \Theta + c$ and $\hat \tau^{\dag} = \hat \tau$. Also,
let $\hat \sigma^{\dag} = \hat \sigma$. Note that $(\hat \Theta^{\dag}, \hat \tau^{\dag})$ and
$(\hat \sigma^{\dag})^2$ are independent random vectors with
$(n-p) (\hat \sigma^{\dag})^2/(\sigma^{\dag})^2 \sim \chi_{n-p}^2$.
The uncertain prior information may be re-expressed as $\tau^{\dag}=0$.

This re-expressed model and prior information have the same form as
the original model and prior information. Thus the confidence interval
$\big [ \ell(\hat \Theta^{\dag}, \hat \tau, \sigma),
u(\hat \Theta^{\dag}, \hat \tau, \sigma) \big ]$  for $\theta^{\dag}$ must lead to a
confidence interval for $\theta$ that is identical to \eqref{A.1}. This implies that
$\ell(\hat \Theta, \hat \tau, \hat \sigma) = \hat \Theta + \tilde \ell(\hat \tau, \hat \sigma)$
and $u(\hat \Theta, \hat \tau, \hat \sigma) = \hat \Theta + \tilde u(\hat \tau, \hat \sigma)$,
%
%
where $\tilde \ell: \mathbb{R} \times [0,\infty) \rightarrow \mathbb{R}$
and $\tilde u: \mathbb{R} \times [0,\infty) \rightarrow \mathbb{R}$.

\medskip

\noindent {\bf Invariance Argument 2}

\noindent Let $c$ be a positive number. The model for the reduced data may be re-expressed
\begin{equation*}
\left[\begin{matrix} \hat \Theta^{\dag}\\ \hat \tau^{\dag} \end{matrix}
\right] \sim N \left ( \left[\begin{matrix} \theta^{\dag} \\
\tau^{\dag} \end{matrix}
\right], (\sigma^{\dag})^2 V \right ).
\end{equation*}
where $\theta^{\dag} = c \, \theta$, $\tau^{\dag} = c \, \tau$, $\sigma^{\dag} = c \, \sigma$,
$\hat \Theta^{\dag} = c \, \hat \Theta$ and $\hat \tau^{\dag} = c \, \hat \tau$. Also,
let $\hat \sigma^{\dag} = c \, \hat \sigma$. Note that $(\hat \Theta^{\dag}, \hat \tau^{\dag})$ and
$(\hat \sigma^{\dag})^2$ are independent random vectors with
$(n-p) (\hat \sigma^{\dag})^2/(\sigma^{\dag})^2 \sim \chi_{n-p}^2$.
The uncertain prior information may be re-expressed as $\tau^{\dag}=0$.

This re-expressed model and prior information have the same form as
the original model and prior information. Thus the confidence interval
$\big [ \hat \Theta^{\dag} + \tilde \ell(\hat \tau^{\dag}, \hat \sigma^{\dag}),
\hat \Theta^{\dag} + \tilde u(\hat \tau^{\dag}, \hat \sigma^{\dag})  \big ]$  for $\theta^{\dag}$ must lead to a
confidence interval for $\theta$ that is identical to
$\big [ \hat \Theta + \tilde \ell(\hat \tau, \hat \sigma),
\hat \Theta + \tilde u(\hat \tau, \hat \sigma)  \big ]$ for $\theta$.
This implies that
$\ell(\hat \Theta, \hat \tau, \hat \sigma)
= \hat \Theta - \tilde b (\hat \tau/\hat \sigma) \hat \sigma
- \tilde s (\hat \tau/\hat \sigma) \hat \sigma$
and
$u(\hat \Theta, \hat \tau, \hat \sigma)
= \hat \Theta - \tilde b (\hat \tau/\hat \sigma) \hat \sigma
+ \tilde s (\hat \tau/\hat \sigma) \hat \sigma$,
%
%
where $\tilde b: \mathbb{R} \rightarrow \mathbb{R}$ and $\tilde s: \mathbb{R} \rightarrow [0,\infty)$.

\medskip

\noindent {\bf Invariance Argument 3}

\noindent The model for the reduced data may be re-expressed
\begin{equation*}
\left[\begin{matrix} \hat \Theta^{\dag}\\ \hat \tau^{\dag} \end{matrix}
\right] \sim N \left ( \left[\begin{matrix} \theta^{\dag} \\
\tau^{\dag} \end{matrix}
\right], (\sigma^{\dag})^2 V \right ).
\end{equation*}
where $\theta^{\dag} = - \theta$, $\tau^{\dag} = - \tau$, $\sigma^{\dag} = \sigma$,
$\hat \Theta^{\dag} = - \hat \Theta$ and $\hat \tau^{\dag} = - \hat \tau$. Also,
let $\hat \sigma^{\dag} = \hat \sigma$. Note that $(\hat \Theta^{\dag}, \hat \tau^{\dag})$ and
$(\hat \sigma^{\dag})^2$ are independent random vectors with
$(n-p) (\hat \sigma^{\dag})^2/(\sigma^{\dag})^2 \sim \chi_{n-p}^2$.
The uncertain prior information may be re-expressed as $\tau^{\dag}=0$.

This re-expressed model and prior information have the same form as
the original model and prior information. Thus the confidence interval
\begin{equation*}
\left [\hat \Theta^{\dag} - \tilde b \left(\frac{\hat \tau^{\dag}}{\hat \sigma^{\dag}}\right) \hat \sigma^{\dag}
- \tilde s \left(\frac{\hat \tau^{\dag}}{\hat \sigma^{\dag}}\right) \hat \sigma^{\dag},
\, \hat \Theta^{\dag} - \tilde b \left(\frac{\hat \tau^{\dag}}{\hat \sigma^{\dag}}\right) \hat \sigma^{\dag}
+ \tilde s \left(\frac{\hat \tau^{\dag}}{\hat \sigma^{\dag}}\right) \hat \sigma^{\dag} \right]
\end{equation*}
for $\theta^{\dag}$ must lead to a
confidence interval for $\theta$ that is identical to the confidence interval
\begin{equation*}
\left [\hat \Theta - \tilde b \left(\frac{\hat \tau}{\hat \sigma}\right) \hat \sigma
- \tilde s \left(\frac{|\hat \tau|}{\hat \sigma}\right) \hat \sigma, \, \hat \Theta - \tilde b \left(\frac{\hat \tau}{\hat \sigma}\right) \hat \sigma
+ \tilde s \left(\frac{|\hat \tau|}{\hat \sigma}\right) \hat \sigma \right]
\end{equation*}
for $\theta$.
This implies that
$\tilde b$ is an odd function and $\tilde s: [0,\infty) \rightarrow [0,\infty)$.

Now define the functions
$b(x) = (1/\sqrt{v_{11}})  \, \tilde{b} \big(\sqrt{v_{22}}x \big)$ for all $x \in \mathbb{R}$ and
$s(x) = (1/\sqrt{v_{11}}) \, \tilde s \big(\sqrt{v_{22}}x \big)$ for all $x \ge 0$.
%
%
%
%
Since $\tilde b$ is constrained to be an odd function, $b$ is also an
odd function. Also, since $\tilde s: [0, \infty) \rightarrow [0, \infty)$,
$s: [0, \infty) \rightarrow [0, \infty)$. The confidence interval \eqref {A.1} is therefore equal to $J(b,s)$
where $b: \mathbb{R} \rightarrow \mathbb{R}$ is an odd function and $s: [0, \infty) \rightarrow [0, \infty)$.

\smallskip

\bigskip

\noindent {\large{\bf {Appendix B. Proof of Theorem 1}}}

\medskip



\noindent {\bf Proof of part (a).}

\smallskip

\noindent The random vectors $(G,H)$ and $W$ are independent. It follows from \eqref{model_G_H}
that the probability density function of $H$, evaluated at $h$, is $\phi(h-\gamma)$. Thus
\begin{equation}
\tag{B.1}
\label{B.1}
 c(\gamma;b,s,\rho)
 = \int_0^{\infty} \int_{-\infty}^{\infty} \int_{\ell(h,w)}^{u(h,w)}
f_{G|H}(g|h) \, dg \, \phi(h-\gamma)\,  dh \, f_W(w) \, dw
\end{equation}
where $f_W$ denotes the probability density function of $W$ and $f_{G|H}(g|h)$ denotes the probability density function of $G$ conditional on $H=h$, evaluated at $g$.
The probability distribution of $G$ conditional on $H=h$ is
$N \big( \rho (h - \gamma), 1 - \rho^2 \big )$. Thus the right hand side of
\eqref{B.1} is equal to
\begin{equation}
\tag{B.2}
\label{B.2}
\int_0^{\infty} \int_{-\infty}^{\infty} k(h,w, \gamma, \rho)
\, \phi(h-\gamma)\,  dh \, f_W(w) \, dw
\end{equation}
The standard $1-\alpha$ confidence interval $I$ has coverage probability $1-\alpha$.
Hence
\begin{equation}
\tag{B.3}
\label{B.3}
1-\alpha = \int_0^{\infty} \int_{-\infty}^{\infty} k^{\dag}(h,w, \gamma, \rho)
\, \phi(h-\gamma)\,  dh \, f_W(w) \, dw.
\end{equation}
Subtracting \eqref{B.3} from \eqref{B.2}  and noting that
$b(x)=0$ for all $|x| \ge d$ and $s(x)=t_{n-p,1-\frac{\alpha}{2}}$ for all $x \ge d$, we find that
\begin{equation*}
 c(\gamma;b,s,\rho)
 = (1-\alpha) +
\int_0^{\infty} \int_{-dw}^{dw} \big( k(h,w, \gamma, \rho) - k^{\dag}(h,w, \gamma, \rho) \big)
\, \phi(h-\gamma)\,  dh \, f_W(w) \, dw.
\end{equation*}
Changing the variable of integration from $h$ to $x=h/w$ in the inner integral, we obtain \eqref{cov_prob}.
Using the fact that
\begin{equation*}
\left[\begin{matrix} -G\\ -H \end{matrix}
\right] \sim N \left ( \left[\begin{matrix} 0 \\ -\gamma \end{matrix}
\right], \left[\begin{matrix} 1 \quad \rho\\ \rho \quad 1 \end{matrix}
\right] \right ),
\end{equation*}
it may be shown that $P(\theta \in J(b,s))$ is an even function of $\gamma$.

\smallskip

\noindent{\bf Proof of part (b).}

\smallskip

\noindent The random variables $H$ and $W$ are independent. It follows from \eqref{model_G_H}
that the probability density function of $H$, evaluated at $h$, is $\phi(h-\gamma)$. Thus
\begin{equation}
\tag{B.4}
\label{B.4}
 e(\gamma;s) =
 \frac{1} {t_{n-p, 1 - \frac{\alpha}{2}} \, E(W)}
 \int^{\infty}_0 \int^{\infty}_{-\infty} s \left (\frac{|h|}{w}
\right) \phi(h-\gamma) \, dh \,  w \,  f_W(w)  \, dw
\end{equation}
where $f_W$ denotes the probability density function of $W$. Obviously,
\begin{equation}
\tag{B.5}
\label{B.5}
 1 = \frac{1} {t_{n-p, 1 - \frac{\alpha}{2}} \, E(W)}
 \int^{\infty}_0 \int^{\infty}_{-\infty} t_{n-p, 1 - \frac{\alpha}{2}} \, \phi(h-\gamma)
 \, dh \,  w \,  f_W(w)  \, dw.
\end{equation}
Note that $s(x)=t_{n-p,1-\frac{\alpha}{2}}$ for all $x \ge d$. Subtracting \eqref{B.5}
from \eqref{B.4} we therefore obtain
\begin{equation*}
 e(\gamma;s) = 1 +
 \frac{1} {t_{n-p, 1 - \frac{\alpha}{2}} \, E(W)}
 \int^{\infty}_0 \int^{d w}_{- d w} \left (s \left (\frac{|h|}{w}
\right) - t_{n-p,1-\frac{\alpha}{2}} \right ) \phi(h-\gamma) \, dh \,  w \,  f_W(w)  \, dw.
\end{equation*}
Changing the variable of integration in the inner integral from $h$ to $x = h/w$, we obtain
\eqref{comp_conv_e}.

\baselineskip=21pt

\bigskip

\noindent {\bf References}

\medskip

\rf Bickel, P.J., 1983. Minimax estimation of the mean of a normal distribution subject
to doing well at a point. In:  Rizvi, M.H.,  Rustagi, J.S., Siegmund, D., (Eds), Recent Advances in Statistics, Academic
Press, New York, 511--528.

\smallskip

\rf Bickel, P.J., 1984. Parametric robustness: small biases can be worthwhile.
Annals of Statistics 12, 864--879.

\smallskip

\rf Bohrer, R., Sheft, J., 1979. Misclassification probabilities in $2^3$ factorial experiments.
Journal of Statistical Planning and Inference 3, 79--84.

\smallskip

\rf Box, G.E.P., Connor, L.R., Cousins, W.R., Davies, O.L., Hinsworth, F.R., Sillitto, G.P., 1963. The Design and Analysis
of Industrial Experiments, 2nd edition, reprinted. Oliver and Boyd, London.



\smallskip

\rf Buring, J.E., Hennekens, C.H., 1990. Cost and efficiency in clinical trials:
the U.S. Physicians' Health Study. Statistics in Medicine 9, 29--33.

\smallskip

\rf Casella, G., Berger, R. L., 2002. Statistical
Inference, 2nd ed. Duxbury, Pacific Grove, California.

\smallskip

\rf Dub\'e, M.A., Penlidis, A., Reilly, P.M., 1996. A systematic approach to the study of multicomponent polymerization
kinetics: the butyl acrylate/methyl methacrylate/vinyl acetate example. IV. Optimal Bayesian design of emulsion
terpolymerization experiments in a pilot plant reactor. Journal of Polymer Science: Part A: Polymer Chemistry 34, 811--831.

\smallskip

\rf Fabian, V., 1991. On the problem of interactions in the analysis of variance.
Journal of the American Statistical Association 86, 362--373.

\smallskip

\rf Farchione, D., Kabaila, P., 2008. Confidence intervals for the normal mean utilizing prior information.
Statistics \& Probability Letters, 78, 1094--1100.

\smallskip

\rf Giri, K., 2008. Confidence intervals in regression utilizing prior information.
Unpublished PhD thesis, August 2008, Department of Mathematics and Statistics, La Trobe University.

\smallskip

\rf Giri, K., Kabaila, P., 2008. The coverage probability of confidence intervals in $2^r$ factorial
experiments after preliminary hypothesis testing.
Australian \& New
Zealand Journal of Statistics, 50, 69--79.



\smallskip

\rf Hodges, J.L., Lehmann, E.L., 1952. The use of previous
experience in reaching statistical decisions. Annals of Mathematical
Statistics 23, 396--407.

\smallskip

\rf Hung, H.M.J., Chi, G.Y.H., O'Neill, R.T., 1995. Efficiency evaluation of monotherapies in two-by-two factorial
trials. Biometrics 51, 1483--1493.

\smallskip

\rf Joshi, V.M., 1969. Admissibility of the usual confidence sets for the mean of a univariate or bivariate normal
population. Annals of Mathematical Statistics, 40, 1042--1067.

\smallskip

\rf Kabaila, P., 1998. Valid confidence intervals in
regression after variable selection. Econometric Theory 14,
463--482.

\smallskip

\rf Kabaila, P., 2005a. On the coverage probability of confidence intervals
in regression after variable selection. Australian \& New Zealand Journal of Statistics
47, 549--562.

\smallskip

\rf Kabaila, P., 2005b. Assessment of a preliminary F-test solution to the Behrens-Fisher
problem. Communications in Statistics - Theory and Methods 34, 291--302.

\smallskip

\rf Kabaila, P., Giri, K., 2007a. Large sample confidence intervals in regression utilizing prior information.
Technical Report No. 2007-1, Jan 2007, Department of Mathematics and Statistics, La Trobe University.

\smallskip

\rf Kabaila, P., Giri, K., 2007b. Confidence intervals in regression utilizing prior information.
arXiv:0711.3236v1.

\smallskip



\rf Kabaila, P., Giri, K., 2009a. Upper bounds on the minimum coverage probability of confidence
intervals in regression after variable selection. To appear in Australian \& New Zealand Journal of Statistics.

\smallskip

\rf Kabaila, P., Giri, K., 2009b. Large-sample confidence intervals for the treatment difference
in a two-period crossover trial, utilizing prior information. Statistics \& Probability
Letters 79, 652--658.

\smallskip

\rf Kabaila, P., Leeb, H., 2006. On the large-sample minimum coverage
probability of confidence intervals after
model selection. Journal of the American Statistical Association 101, 619--629.

\smallskip

\rf Kabaila, P., Tuck, J., 2008. Confidence intervals utilizing prior information in the Behrens-Fisher problem.
Australian \& New Zealand Journal of Statistics 50, 309--328.



\smallskip

\rf Neyman, J., 1935. Comments on `Complex experiments' by F. Yates.
Supplement to the Journal of the Royal Statistical Society 2, 181--247.

\smallskip

\rf Ng, T-H., 1994. The impact of a preliminary test for interaction in a $2 \times 2$ factorial trial.
Communications in Statistics: Theory and Methods 23, 435--450.

\smallskip

\rf Pratt, J.W., 1961. Length of confidence intervals.
Journal of the American Statistical Association 56, 549--657.

\smallskip

\rf Shaffer, J.P., 1991. Probability of directional errors with disordinal (qualitative) interaction.
Psychometrika 56, 29--38.

\smallskip

\rf Stampfer, M.J., Buring, J.E., Willett, W., Rosner, B., Eberlain, K., Hennekens, C.H., 1985. The $2 \times 2$ factorial
design: its application to a randomized trial of aspirin and carotene in U.S. physicians. Statistics in Medicine
4, 111---116.

\smallskip

\rf Steering Committee of the Physicians' Health Study Research Group: Belanger, C., Buring, J.E., Cook, N., Eberlain, K.,
Goldhaber, S.Z., Gordon, D., Hennekens, C.H. (chairman), Mayrent, S.L., Peto, R., Rosner, B., Stampfer, M.J.,
Stubblefield, F., Willett, W., 1988. Preliminary report: findings from the aspirin component of the ongoing
Physicians' Health Study. New England Journal of Medicine 318, 262--264.

\smallskip

\rf Traxler, R.H., 1976. A snag in the history of factorial experiments.
In: Owen, D.B. (Ed), On the History of Statistics and Probability, Marcel Dekker, New York, 281--295.

\smallskip

\rf Tuck, J., 2006. Confidence intervals incorporating prior
information. Unpublished PhD thesis, August 2006, Department of Mathematics and
Statistics, La Trobe University.

\end{document}